\documentclass[12pt]{article}
\usepackage{amsmath,amssymb,amscd,cite}
\usepackage{hyperref}
\usepackage[pagewise]{lineno}
\usepackage[english]{babel}
\usepackage{mathrsfs}
\newtheorem{thm}{Theorem}[section]
\newtheorem{prop}[thm]{Proposition}
\newtheorem{lem}[thm]{Lemma}
\newtheorem{cor}[thm]{Corollary}

\newcommand{\pf}{{\bf Proof. \ }}
\newcommand{\qed}{\hfill $\blacksquare$ \\}

\newtheorem{ex}[thm]{Example}

\newcommand{\m}{mod}

\date{\today}
\begin{document}

\title{Linear Codes over $\mathfrak{R}^{s,m}=\sum\limits_{\varsigma=1}^{m} v_{m}^{\varsigma-1}\mathcal{A}_{m-1}$, with $v_{m}^{m}=v_{m}$}
	\author{Mouna. Malki \thanks{Faculty of Mathematics and informatics, Department of Mathematics, Mostefa Ben Boula\"{i}d University, Batna 2,
			Batna, Algeria. E-mail: karima.chatouh@gmail.com} and Karima. Chatouh \thanks{Faculty of Mathematics and informatics, Department of Mathematics, Mostefa Ben Boula\"{i}d University, Batna 2,
			Batna, Algeria. E-mail: malki\_mouna@yahoo.com}}
	\maketitle
	
	\begin{abstract}
	The main objective of this paper is to extend the previously defined code family over the ring $\mathfrak{R}=\sum\limits_{s=0}^{4} v_{5}^{s} \mathcal{A}_{4}$ to $\mathfrak{R}^{s,m}=\sum\limits_{\varsigma=1}^{m} v_{m}^{\varsigma-1}\mathcal{A}_{m-1}$, and propose an expanded framework for its implementation in coding theory, and to derive additional properties from this generalized code family, including the construction of cyclic and quasi-cyclic codes. Furthermore, we will present specific applications of this extended code family.
	\end{abstract}
	
	\section{Introduction}\label{sec 1}

	Bustomi et al. \cite{article8} extended the work of Li et al. \cite{article16} by generalizing the ring $\mathbb{Z}_{4} + u\mathbb{Z}_{4} + v\mathbb{Z}_{4} + uv\mathbb{Z}_{4}$ to include the additional elements $w\mathbb{Z}_{4}$, $uw\mathbb{Z}{4}$, $vw\mathbb{Z}_{4}$, and $uvw\mathbb{Z}_{4}$ while imposing the conditions $u_{2} = u$, $v_{2} = v$, $w_{2} = w$, $uv = vu$, $uw = wu$, and $vw = wv$. The authors investigated linear codes over this ring and derived some related properties.
	
	 Furthermore, Ndiaye and Gueye \cite{Nadiaye}, investigated cyclic codes over the ring $F_{p}^{k} +vF_{p}^{k} +v^{2}F_{p}^{k} +\dots+v^{r}F_{p}^{k}$ , where $v^{r+1} = v$, $p$ a prime number, $r\ge 1$ and $gcd(r; p) = 1$. Ndiaye and Gueye \cite{Nadiaye} generalized the results of \cite{article14}, namely; they showed that the codes are principally generated, they gave
	a generator polynomial for these codes and they showed that the idempotent depends on the idempotents of the above ring \cite{article12}.\\
	
	In our study, we expanded the ring that was investigated in Bustomi et al \cite{article8} to include the ring $\mathcal{A}_{4}$. We replaced $F_{p}^{k}$ in the ring $F_{p}^{k} +vF_{p}^{k} +v^{2}F_{p}^{k} +\dots+v^{r}F_{p}^{k}$ with $\mathcal{A}_{4}$, which was previously examined by Ndiaye et al \cite{Nadiaye}. Drawing inspiration from \cite{Mouna}, we combined these two rings to investigate the generalization of the ring $\mathfrak{R}=\sum\limits_{s=0}^{4} v_{5}^{s} \mathcal{A}_{4}$, where $v_{5}^{5}=v_{5}$. Our approach led us to study the properties of the generalized ring $\mathfrak{R}^{s,m}=\sum\limits_{\varsigma=1}^{m} v_{m}^{\varsigma-1}\mathcal{A}_{m-1}$ with $v_{m}^{m}=v_{m}$. In the following sections, we will delve into the new properties that we have uncovered using this approach.\\

	The inspiration for this paper comes from the findings in and  \cite{Mouna}. Our main focus is on studying the ring $\mathfrak{R}^{s,m}=\sum\limits_{\varsigma=1}^{m} v_{m}^{\varsigma-1}\mathcal{A}_{m-1}$ with $v_{m}^{m}=v_{m}$ which is a composite of two rings $ \mathbb{Z}_{4^{s}}$ and $\mathcal{A}_{m-1} $. This represents a novel approach to constructing rings, and through our approach, we have uncovered new results and findings about the idempotent elements. Moreover, we have developed new structural properties of linear codes over $\mathfrak{R}^{s,m}$, using the Chinese Remainder Theorem and our novel approach, which is based on the composite ring. We have also constructed generator matrices with the new idempotents and Gray images. Our findings and ring construction are demonstrated through three applications. \\
	
	The paper is structured as follows: Firstly, Section \ref{2} presents the preliminaries. Following that, Section 3 defines a composite of two Gray maps from $\mathfrak{R}^{s,m}$ to $\mathbb{Z}_{4^{s}}^{3\times2^{m-1}}$. and establishes a linear code over $\mathfrak{R}^{s,m}$, along with generator matrices and Gray images. while Section 3. In Section 4, we study the quasi-cyclic codes from simplex codes over $\mathfrak{R}^{s,m}$. Furthermore, we showcase some applications and examples.

	\section{Preliminaries}\label{2}
	
	The focus of this study is on constructing linear codes over the ring $\mathfrak{R}^{s,m}$ by utilizing linear codes over the ring $\mathcal{A}_{m-1}$.\\
	
	The definition of $\mathfrak{R}^{s,m}$ is given in \cite{Mouna}, but in a more general form. We have

	The ring $\mathfrak{R}^{s,m}=\mathcal{A}_{m-1}+v_{m}\mathcal{A}_{m-1}+\ldots+v_{m}^{m-1}\mathcal{A}_{m-1}$, where $\mathcal{A}_{m-1}=\mathbb{Z}_{4^{s}}[v_{1},v_{2},\ldots,v_{m-1}] \diagup \langle v_{i}^{2}=v_{i}, v_{i}v_{j}=v_{j}v_{i}\rangle $,
	$i,j=\overline{1,m-1}$, $i \neq j$ and $s \geq 1$ is a commutative Frobenius ring, and  $\vert \mathfrak{R}^{s,m}\vert= 4^{ms2^{m-1}}$. 
	We can also regard $\mathfrak{R}^{s,m}$, as a quotient ring of a polynomial ring over $\mathbb{Z}_{4^{s}}[v_{1},v_{2}\ldots v_{m}]/<v^{2}_{i}-v_{i},v_{m}^{m}-v_{m}>$, for $i=\overline{1,m-1}$.\\

	Based on the definition of orthogonal, non-zero idempotents in a commutative ring $\mathfrak{R}^{s,m}$, we can state the following proposition.
	\begin{prop}\label{prop1}
		Let $\kappa_{1}, \kappa_{2}$ and $\kappa_{3}$ be three elements of $\mathfrak{R}^{s,m}$. Then $\kappa_{i}$, for $i=\overline{1,3}$ are orthogonal non-zero idempotents verifying the Pierce conditions over $\mathfrak{R}^{s,m},$ where
		\[\kappa_{1}=\prod_{i=1}^{m-1} (1-v_{i}) \dfrac{1}{m-1}[v_{m}+v_{m}^{2}+\ldots+v_{m}^{m-1}],\]
		
		\[\kappa_{2}=\prod_{i=1}^{m-1} (1-v_{i}) \dfrac{-1}{m-1}[v_{m}+v_{m}^{2}+\ldots-(m-2)v_{m}^{m-1}],\]
		and
		\[\kappa_{3}=1-\prod_{i=1}^{m-1} (1-v_{i})v_{m}^{m-1}. \]
	\end{prop}
	\pf
	By [\cite{Mouna}, Proposition 1], we obtain
		\begin{enumerate}
		\item[1.] \begin{eqnarray*}
			\sum\limits_{i=1}^{3}\kappa_{i} &=& 1+\dfrac{1}{m-1}\prod_{i=1}^{m-1}(1-v_{i})\left[ v_{m}+v_{m}^{2}+\ldots
			-v_{m}-v_{m}^{2}- \ldots \right. \\
			& & \left. +(m-2)v_{m}^{m-1}-(m-1)v_{m}^{m-1}\right] \\
			&=& 1+ \dfrac{1}{m-1}\prod_{i=1}^{m-1}(1-v_{i})\left[  v_{m}^{m-1}+(m-2)v_{m}^{m-1}-(m-1)v_{m}^{m-1}\right] \\
			&=& 1  
		\end{eqnarray*}
		\item[2.] We show that $\kappa_{i}^{2}=\kappa_{i}$, for $i=\overline{1,3}$, we have
		\begin{eqnarray*}
			\kappa_{1}^{2}&=&\left[ \prod_{i=1}^{m-1} (1-v_{i})\right] ^{2}\dfrac{1}{(m-1)^{2}}\left( v_{m}+v_{m}^{2}+\ldots+v_{m}^{m-1}\right) \left( v_{m}+v_{m}^{2}+\ldots+v_{m}^{m-1}\right) \\
			&=&\prod_{i=1}^{m-1}\left( 1-v_{i})\dfrac{1}{(m-1)^{2}}(v^{2}_{m}+v^{3}_{m}+v^{4}_{m}+\ldots+v^{m-1}_{m}+v^{m}_{m}+v^{3}_{m}+v^{4}_{m}+\ldots\right. \\
			& &\left. +v^{m-1}_{m}+v^{m}_{m}+v^{m+1}_{m}+\ldots+v^{m}_{m}+v^{m+1}_{m}+v^{m+2}_{m}+\ldots + v^{2(m-1)}_{m}\right) \\
			&=& \prod_{i=1}^{m-1}(1-v_{i})\dfrac{1}{(m-1)^{2}}\left( v_{m}^{2}+2v_{m}^{3}+3v_{m}^{4}+\ldots+(j-1)v_{m}^{j}+\right. \\
			& &\ldots+(m-2)v_{m}^{m-1}+(m-1)v_{m}^{m}+(m-2)v_{m}^{m+1}+(m-3)v_{m}^{m+2}+\\
			& &\left. \ldots+(m-j)v_{m}^{m+j-1}+...+v_{m}^{m-1}\right) \\
			&=&\prod_{i=1}^{m-1}(1-v_{i})\dfrac{1}{(m-1)^{2}}\left[ (m-1)v_{m}+(m-1)v_{m}^{2}+(m-1)v_{m}^{3}+(m-1)v_{m}^{4}+\right. \\
			& &\left. \ldots+(m-1)v_{m}^{j}+
			\ldots+(m-1)v_{m}^{m-1}+(m-1)v_{m}^{m}\right] \\
			&=&\kappa_{1}.
		\end{eqnarray*}
		Similar arguments for $\kappa_{2}$.
		
		Likewise for $\kappa_{3}$, we have
		\begin{eqnarray*}
			\kappa_{3}^{2}&=& \left[1- \prod_{i=1}^{m-1} (1-v_{i})v_{m}^{(m-1)}\right] ^{2}\\
			&=& 1-2\prod_{i=1}^{m-1} (1-v_{i})v_{m}^{m-1}+\prod_{i=1}^{m-1} (1-v_{i})v_{m}^{m-1}\\
			&=& 1-\prod_{i=1}^{m-1} (1-v_{i})v_{m}^{m-1}\\
			&=& \kappa_{3}.
		\end{eqnarray*}
		\item[3.] Now, we are verifying $\kappa_{i}\cdot \kappa_{j}=0$, for $1 \leq i \neq j \leq 3$, 
		\begin{eqnarray*}
			\kappa_{1} \cdot \kappa_{2}&=&
			\prod_{i=1}^{m-1} (1-v_{i})\dfrac{1}{m-1}(v_{m}+v_{m}^{2}+ \ldots +v_{m}^{m-1})\\
			&&\prod_{i=1}^{m-1} (1-v_{i})(\dfrac{-1}{m-1})(v_{m}+v_{m}^{2}+\ldots -(m-2)v_{m}^{m-1})\\
			&=&\kappa_{1}\left[ -\kappa_{1}+\prod_{i=1}^{m-1} (1-v_{i})v_{m}^{m-1}\right] \\
			&=& -\kappa_{1}^{2}+\kappa_{1}\\
			&=& 0.\\
		\end{eqnarray*}
		Moreover,
		\begin{eqnarray*}
			\kappa_{2} \cdot \kappa_{3}&=& \prod_{i=1}^{m-1} (1-v_{i})(\dfrac{-1}{m-1})(v_{m}+v_{m}^{2}+\ldots-(m-2)v_{m}^{m-1})
			\left[ 1-\prod_{i=1}^{m-1} (1-v_{i})v_{m}^{m-1}\right] \\
			&=& \kappa_{2}-\left[ \prod_{i=1}^{m-1} (1-v_{i})\right] ^{2}(\dfrac{1}{m-1})(-v_{m}^{m}-v_{m}^{m+1}-\ldots+(m-2)v_{m}^{2(m-1)})\\
			&=& \kappa_{1}-\kappa_{1}=0.
		\end{eqnarray*}
	\end{enumerate} 
	Similar arguments hold for $\kappa_{1} \cdot \kappa_{3}=0.$                             
	\qed
	According to Proposition \ref{prop1}, we decompose $\mathfrak{R}^{s,m}$ as the direct sum of three components
	\begin{equation}\label{eq1}
	\mathfrak{R}^{s,m}= \kappa_{1}\mathfrak{R}^{s,m}\oplus \kappa_{2}\mathfrak{R}^{s,m}\oplus \kappa_{3}\mathfrak{R}^{s,m}.
	\end{equation}
Additionally we can express the ring $\mathfrak{R}^{s,m}$ by
	\begin{equation}\label{eq2}
	\mathfrak{R}^{s,m}= \kappa_{1}\mathcal{A}_{m-1}\oplus \kappa_{2}\mathcal{A}_{m-1}\oplus \kappa_{3}\mathcal{A}_{m-1}.
	\end{equation}
	Consider the following idempotent elements of $\mathcal{A}_{m-1},$
	\begin{eqnarray*}
	\eta_{0} & = & \prod\limits_{i=1}^{m-1}\left( 1-v_{i}\right)\\
	\eta_{i} & = & v_{i}\prod \limits_{j=1}^{m-1}\left( 1-v_{j}\right), \  i \neq j,\  1\leq i\leq m-1, \  \vert \eta_{i} \vert =\left( \begin{array}{c}
	1\\m-1
	\end{array}\right) \\
	\eta_{ij} & = & v_{i}v_{j}\prod \limits_{k=1}^{m-1}\left( 1-v_{k}\right), \  i \neq j \neq k, \  1 \leq i,j\leq m-1,\  \vert \eta_{ij} \vert =\left( \begin{array}{c}
	2\\m-1
	\end{array}\right)\\
	& \vdots & \\
	\eta_{2^{m-1}-1} & = & \prod\limits_{k=1}^{m-1}v_{k},
	\end{eqnarray*}
	where
	\begin{eqnarray*}
	\vert \eta_{0} \vert+ \vert \eta_{i} \vert+ \vert \eta_{ij}\vert + \ldots + \vert \eta_{2^{m-1}-1}\vert &=&1 +\left( \begin{array}{c}
	1\\m-1
	\end{array}\right)+\left( \begin{array}{c}
	2\\m-1
	\end{array}\right) +\ldots+1\\
	&=&2^{m-1}.
	\end{eqnarray*}
	The upper elements are also pairwise orthogonal, since $\eta_{\iota}\eta_{\zeta}=0$ and
$\sum\limits_{\iota=0}^{2^{m-1}-1}\eta_{\iota}=1$, for $\iota\neq \zeta$ and also $\iota,\zeta  \in \left\lbrace 0, i ,ij, \ldots, 2^{m-1}-1 \right\rbrace $. Consequently, by the Chinese Remainder Theorem, we obtain
\begin{equation}
\mathcal{A}_{m-1}= \eta_{0} \mathcal{A}_{m-1} \oplus \eta_{i} \mathcal{A}_{m-1} \oplus \ldots\oplus \eta_{2^{m-1}-1} \mathcal{A}_{m-1},
\end{equation}
and
	\begin{equation}
	\mathcal{A}_{m-1}= \eta_{0} \mathbb{Z}_{4^{s}} \oplus \eta_{i} \mathbb{Z}_{4^{s}} \oplus \ldots\oplus \eta_{2^{m-1}-1} \mathbb{Z}_{4^{s}}.
	\end{equation}
	\begin{lem} \cite{Mouna,Nadiaye}
	The element $ \tau=\kappa_{1}\eta_{\iota} + \kappa_{2}\eta_{\zeta} +\kappa_{3} \eta_{\lambda} $ is an idempotent in $\mathfrak{R}^{s,m}[x]/( x^{n}-1)  $ if only if $\eta_{\iota},\eta_{\zeta}$ and $\eta_{\lambda}$ are an idempotents in $\mathcal{A}_{m-1}[x]/( x^{n}-1),$ for all $\iota,\zeta$ and $\lambda$	  $ \in \left\lbrace 0, i ,ij, \ldots, 2^{m-1}-1 \right\rbrace.$ 
\end{lem}
\pf
If  $ \tau=\kappa_{1}\eta_{\iota} + \kappa_{2}\eta_{\zeta} +\kappa_{3} \eta_{\lambda} $ is an idempotent in $\mathfrak{R}^{s,m}[x]/( x^{n}-1)  $, then
\begin{eqnarray*}
\tau^{2}=
	[ \kappa_{1}\eta_{\iota} + \kappa_{2}\eta_{\zeta} +\kappa_{3} \eta_{\lambda} ]^{2}&=& \kappa_{1}\eta_{\iota}^{2} + \kappa_{2}\eta_{\zeta}^{2} +\kappa_{3} \eta_{\lambda}^{2}\\
	&=& \kappa_{1}\eta_{\iota} + \kappa_{2}\eta_{\zeta} +\kappa_{3} \eta_{\lambda} \\
		&=& \tau,
\end{eqnarray*}
that means $\eta_{\iota}^{2}=\eta_{\iota}$,$\eta_{\zeta}^{2}=\eta_{\zeta} $ and $\eta_{\lambda}^{2}=\eta_{\lambda}$ .

Conversely if $\eta_{\iota},\eta_{\zeta}$and $\eta_{\lambda}$ are an idempotents in $\mathcal{A}_{m-1}[x]/( x^{n}-1),$ for all $\iota,\zeta$ and $\lambda$ $ \in \left\lbrace 0, i ,ij, \ldots, 2^{m-1}-1 \right\rbrace$, then 
\begin{eqnarray*}
			[ \kappa_{1}\eta_{\iota} + \kappa_{2}\eta_{\zeta} +\kappa_{3} \eta_{\lambda} ]^{2}=	 [ \kappa_{1}\eta_{\iota} + \kappa_{2}\eta_{\zeta} +\kappa_{3} \eta_{\lambda} ]. 
\end{eqnarray*}
 Which result $[ \kappa_{1}\eta_{\iota} + \kappa_{2}\eta_{\zeta} +\kappa_{3} \eta_{\lambda} ] $ is an idempotent in $\mathfrak{R}^{s,m}[x]/( x^{n}-1).  $
	\qed	
	\subsection{A Gray Map and Gray Images of Linear Code over $\mathfrak{R}^{s,m}$}
	
According to \cite{Chatouh} and \cite{Nadiaye},\cite{Mouna1} we will build up the Gray map and define the weight in such a way that will provide us a distance preserving isometry.
	From the methods described in \cite{Chatouh,Mouna} and Equation \ref{eq1}, any element of $\mathfrak{R}^{s,m}$ can be expressed as
	$$\mathfrak{R}^{s,m}=\sum_{t=0}^{m-1}v^{t}_{m} \left[ \mathbb{Z}_{4^{s}}+\left( \sum_{i=1}^{m-1}v_{i}\right) \mathbb{Z}_{4^{s}}+\left( \sum_{i=1}^{m-2}v_{i}\sum_{j=i+1}^{m-1}v_{j} \right) \mathbb{Z}_{4^{s}}+\dots +\prod_{k=1}^{m-1}v_{k}\mathbb{Z}_{4^{s}}\right]  $$.\\
	Let $ r\in \mathfrak{R}^{s,m} $, then 	
	\begin{eqnarray*}
	 r&=&\kappa_{1}\sum_{t=0}^{m-1}v^{t}_{m} \left[ a_{0}+\sum\limits_{i=1}^{m-1}v_{i}a_{1}^{i}+\sum\limits_{i=1}^{m-2}\sum\limits_{j=i+1}^{m-1}v_{i}v_{j}a_{2}^{ij}+\ldots+\prod\limits_{k=1}^{m-1}v_{k}a_{2^{(m-1)}-1}\right]\\
	 &+&\kappa_{2}\sum_{t=0}^{m-1}v^{t}_{m}
	  \left[  b_{0}+\sum\limits_{i=1}^{m-1}v_{i}b_{1}^{i}+\sum\limits_{i=1}^{m-2}\sum\limits_{j=i+1}^{m-1}v_{i}v_{j}b_{2}^{ij}+\ldots+\prod\limits_{k=1}^{m-1}v_{k}b_{2^{(m-1)}-1}\right]\\ 
	  &+&\kappa_{3}\sum_{t=0}^{m-1}v^{t}_{m} \left[  c_{0}+\sum\limits_{i=1}^{m-1}v_{i}c_{1}^{i}+\sum\limits_{i=1}^{m-2}\sum\limits_{j=i+1}^{m-1}v_{i}v_{j}c_{2}^{ij}+\ldots+\prod\limits_{k=1}^{m-1}v_{k}c_{2^{(m-1)}-1}\right],
	  \end{eqnarray*}
	   since
	  \begin{eqnarray*}
	r&=&\sum_{t=0}^{m-1}v^{t}_{m}\left[\kappa_{1} a_{0}+\kappa_{1}\left( \sum\limits_{i=1}^{m-1}v_{i}a_{1}^{i}\right) +\kappa_{1}\left( \sum\limits_{i=1}^{m-2}\sum\limits_{j=i+1}^{m-1}v_{i}v_{j}a_{2}^{ij}\right) +\ldots+\kappa_{1}\left( \prod\limits_{k=1}^{m-1}v_{k}a_{2^{(m-1)}-1}\right) \right]\\
	 &+& \sum_{t=0}^{m-1}v^{t}_{m} \left[\kappa_{2} b_{0}+\kappa_{2}\left( \sum\limits_{i=1}^{m-1}v_{i}b_{1}^{i}\right) +\kappa_{2}\left( \sum\limits_{i=1}^{m-2}\sum\limits_{j=i+1}^{m-1}v_{i}v_{j}b_{2}^{ij}\right) +\ldots+\kappa_{2}\left( \prod\limits_{k=1}^{m-1}v_{k}b_{2^{(m-1)}-1}\right) \right]\\ 
	&+& \sum_{t=0}^{m-1}v^{t}_{m} \left[\kappa_{3} c_{0}+\kappa_{3}\left( \sum\limits_{i=1}^{m-1}v_{i}c_{1}^{i}\right) +\kappa_{3}\left( \sum\limits_{i=1}^{m-2}\sum\limits_{j=i+1}^{m-1}v_{i}v_{j}c_{2}^{ij}\right) +\ldots+\kappa_{3}\left( \prod\limits_{k=1}^{m-1}v_{k}c_{2^{(m-1)}-1}\right) \right],
	  \end{eqnarray*}
	   by Equation\ref{eq2}, we have
	 \begin{eqnarray*}
	r &=&\left[ \eta_{0}\left( \kappa_{1}a_{0}+\kappa_{2}b_{0}+\kappa_{3}c_{0}\right) + \sum_{i=1}^{m-1}\eta_{i}v_{i} \left( \kappa_{1}(a_{0}+ a_{1}^{i})+\kappa_{2}(b_{0}+ b_{1}^{i})+\kappa_{3}(c_{0}+ c_{1}^{i})\right) \right. \\
	&+& \dots+
	\left. \eta_{2^{(m-1)}-1} \prod_{k=1}^{m-1}v_{k} \left( \kappa_{1}(a_{0}+a_{1}^{i}+\ldots+a_{2^{(m-1)}-1})+\kappa_{2}(b_{0}+b_{1}^{i}+\ldots+b_{2^{(m-1)}-1})\right.\right.  \\ 
	&& \left. \left. +\kappa_{3}(c_{0}+c_{1}^{i}+\ldots+c_{2^{(m-1)}-1})\right) \right]. 
	\end{eqnarray*}		
	
		The Gray map from $\mathfrak{R}^{s,m}$ to $\mathbb{Z}_{4^{s}}^{3\times2^{m-1}}$, is defined by
		\begin{equation}
		\begin{array}{ccccccc}
		\phi=\Psi_{2}\circ \Psi_{1} & : & \mathfrak{R}^{s,m} & \xrightarrow{\Psi_{1}}  & \mathcal{A}_{m-1}^{3}& \xrightarrow{\Psi_{2}} &\mathbb{Z}_{4^{s}}^{3\times2^{m-1}}\\
		  &   &  r & \mapsto & \Psi_{1}(r) & \mapsto & \phi(r),
		\end{array}
		\end{equation}
		where
		$ \Psi_{1}(r)=\left(a(r),b(r),c(r)\right), $
		
		 whith
		\begin{eqnarray*}
		a(r) &= & \left(a_{0}, a_{1}^{i}, \ldots a_{2^{(m-1)}-1}\right), \\
		b(r) &= & \left(b_{0}, b_{1}^{i}, \ldots b_{2^{(m-1)}-1}\right),\\
		a(r) &= & \left(c_{0}, c_{1}^{i}, \ldots c_{2^{(m-1)}-1}\right).
		\end{eqnarray*}
		In addition,
		$\phi(r)=\Psi_{2}\left(a(r),b(r),c(r)\right)=( a_{0},b_{0},c_{0}, a_{0}+ a_{1}^{i},b_{0}+ b_{1}^{i},c_{0}+ c_{1}^{i},\ldots ,a_{0}+ a_{1}^{i}+\dots+a_{2^{(m-1)}-1}, b_{0}+ b_{1}^{i}+\dots+b_{2^{(m-1)}-1},c_{0}+ c_{1}^{i}+\dots+c_{2^{(m-1)}-1}). $
	
	This Gray map can naturally be extended to $n$-tuples coordinatewise, then we can define
	\begin{equation}
	\begin{array}{ccccc}
	\Phi & : & \mathfrak({R}^{s,m})^{n} & \rightarrow & \mathbb{Z}_{4^{s}}^{n3\times2^{m-1}}\\
	&   &  (r_{1},r_{2}\dots, r_{n}) & \mapsto & \Phi(r_{1},r_{2}\dots, r_{n}),
	\end{array}
	\end{equation}
	
	where 
	\begin{eqnarray}\label{eqq1}
	\Phi(r_{1},r_{2}\dots, r_{n})=\Psi_{2}\left(\left(a(r_{1}),b(r_{1}),c(r_{1})\right),\left(a(r_{2}),b(r_{2}),c(r_{2})\right),\ldots, \left(a(r_{n}),b(r_{n}),c(r_{n})\right) \right).
	\end{eqnarray}
	According to Equation \ref{eqq1}, we have $\Phi(r_{1},r_{2}\dots, r_{n})$ equal to,
	 
$
	(a_{0}^{1},a_{0}^{2},\dots,a_{0}^{n},a_{0}^{1}+ a_{1}^{1i},a_{0}^{2}+ a_{1}^{2i},\dots,a_{0}^{n}+ a_{1}^{ni},\dots,a_{0}^{1}+a_{1}^{1i}+\dots+a_{2^{(m-1)}-1}^{1},
	a_{0}^{2}+a_{1}^{2i}+\dots+a_{2^{(m-1)}-1}^{2},\dots,a_{0}^{n}+a_{1}^{ni}+\dots+a_{2^{(m-1)}-1}^{n};b_{0}^{1},b_{0}^{2},\dots,b_{0}^{n},b_{0}^{1}+ b_{1}^{1i},
	b_{0}^{2}+ b_{1}^{2i},\dots,b_{0}^{n}+ b_{1}^{ni},\dots,b_{0}^{1}+b_{1}^{1i}+\dots+b_{2^{(m-1)}-1}^{1},
	b_{0}^{2}+b_{1}^{2i}+\dots+b_{2^{(m-1)}-1}^{2},\dots,
	b_{0}^{n}+b_{1}^{ni}+\dots+b_{2^{(m-1)}-1}^{n}c_{0}^{1},c_{0}^{2},\dots,c_{0}^{n},c_{0}^{1}+ c_{1}^{1i},c_{0}^{2}+ c_{1}^{2i},\dots,c_{0}^{n}+ c_{1}^{ni},\dots,
	c_{0}^{1}+c_{1}^{1i}+\dots+c_{2^{(m-1)}-1}^{1},
	c_{0}^{2}+c_{1}^{2i}+\dots+c_{2^{(m-1)}-1}^{2},\dots,c_{0}^{n}+c_{1}^{ni}+\dots+c_{2^{(m-1)}-1}^{n}).
$
\begin{ex}
\end{ex}
	\subsection{A Linear Code over $\mathfrak{R}^{s,m}$}
	A code $ \mathfrak{C}$ of length $n$ over $\mathfrak{R}^{s,m}$ is an $\mathfrak{R}^{s,m}$-submodule of $\left( \mathfrak{R}^{s,m}\right) ^{n} $ . 
	We use the notation $\mathfrak{C}^{\perp}$ to denote the dual code of $\mathfrak{C}$, such that
		\begin{eqnarray*}
			\mathfrak{C}^{ 	\perp}=\left\lbrace x\in \mathfrak({R^{s,m}})^{n}:\langle x, y\rangle_{R^{s,m}}=0, y\in \mathfrak{C}\right\rbrace .
		\end{eqnarray*}
We need to define the codes $\mathcal{C},\overline{\mathcal{C}}$ and 
$\overline{\overline{\mathcal{C}}}$, as follows
\begin{eqnarray*}
\mathcal{C}=\left\lbrace a\in \mathcal{A}_{m-1})^{n}  |\exists b,c \in  \mathcal{A}_{m-1})^{n}|\kappa_{1}a+\kappa_{2}b+\kappa_{3}c \in \mathfrak{C} \right\rbrace,\\
\overline{\mathcal{C}}=\left\lbrace b\in \mathcal{A}_{m-1})^{n}  |\exists b,c\in  \mathcal{A}_{m-1})^{n}|\kappa_{1}a+\kappa_{2}b+\kappa_{3}c \in \mathfrak{C} \right\rbrace, \\
\overline{\overline{\mathcal{C}}}=\left\lbrace c\in \mathcal{A}_{m-1})^{n}  |\exists a,b \in  \mathcal{A}_{m-1})^{n}|\kappa_{1}a+\kappa_{2}b+\kappa_{3}c\in \mathfrak{C} \right\rbrace.
\end{eqnarray*}
The next result is useful \cite{article4}, and will be used later in this paper
\begin{thm}
	Let $\mathfrak{C}$ be a linear code of length $n$ over $\mathfrak{R^{s,m}}$. Then we have the following unique decomposition
	\begin{equation}\label{eq5}
	\mathfrak{C}=\kappa_{1}\mathcal{C}\oplus\kappa_{2}\overline{C}\oplus\kappa_{3}\overline{\overline{C}},
	\end{equation}
	and
	\begin{equation}\label{eq6}
	\mathfrak{C}^{ 	\perp}=\kappa_{1}\mathcal{C}^{ 	\perp}\oplus\kappa_{2}\overline{\mathcal{C}^{ 	\perp}}\oplus\kappa_{3}\overline{\overline{\mathcal{C}^{ 	\perp}}}.
	\end{equation}
\end{thm}
Such that
\begin{equation}\label{eq7}
\mathcal{C}=\eta_{0}\mathscr{C}_{0}\oplus\eta_{i}\mathscr{C}_{i}\oplus\dots\oplus\eta_{2^{(m-1)}-1}\mathscr{C}_{2^{m-1}-1},
 \end{equation}
\begin{equation} \label{eq8}
\overline{C}=\eta_{0}\overline{\mathscr{C}}_{0}\oplus\eta_{i}\overline{\mathscr{C}}_{i}\oplus\dots\oplus\eta_{2^{(m-1)}-1}\overline{\mathscr{C}}_{2^{m-1}-1}, 
\end{equation}
\begin{equation}\label{eq9}
\overline{\overline{C}}=\eta_{0}\overline{\overline{\mathscr{C}}}_{0}\oplus\eta_{i}\overline{\overline{\mathscr{C}}}_{i}\oplus\dots\oplus\eta_{2^{(m-1)}-1}\overline{\overline{\mathscr{C}}}_{2^{m-1}-1}. 
\end{equation}
Where
\begin{eqnarray*}
\mathscr{C}_{0}&=&\left\lbrace a_{0}\in \mathbb{Z}^{n}_{4^{s}},\forall a\in \mathcal{C}  \right\rbrace, \\
\mathscr{C}_{i}&=&\left\lbrace a_{0}+a_{1}^{i}\in \mathbb{Z}^{n}_{4^{s}},\forall a\in \mathcal{C}
\right\rbrace  , i \in \left\lbrace 1,\dots,m-1\right\rbrace, \\
& \vdots & \\
	\mathscr{C}_{2^{m-1}-1}&=&\left\lbrace a_{0}+ a_{1}^{i}+\dots+a_{2^{(m-1)}-1}\in \mathbb{Z}^{n}_{4^{s}},\forall a\in \mathcal{C}  \right\rbrace ,\\
\end{eqnarray*}
\begin{eqnarray*}
	\overline{\mathscr{C}}_{0}&=&\left\lbrace b_{0}\in \mathbb{Z}^{n}_{4^{s}},\forall b\in \overline{\mathcal{C}}  \right\rbrace, \\
	\overline{\mathscr{C}}_{i}&=&\left\lbrace b_{0}+b_{1}^{i}\in \mathbb{Z}^{n}_{4^{s}},\forall b\in \overline{\mathcal{C}}
	\right\rbrace  , i \in \left\lbrace 1,\dots,m-1\right\rbrace, \\
	& \vdots & \\
	\overline{\mathscr{C}}_{2^{m-1}-1}&=&\left\lbrace b_{0}+ b_{1}^{i}+\dots+b_{2^{(m-1)}-1}\in \mathbb{Z}^{n}_{4^{s}},\forall b\in \overline{C} \right\rbrace ,\\
\end{eqnarray*}
and
\begin{eqnarray*}
	\overline{\overline{\mathscr{C}}}_{0}&=&\left\lbrace c_{0}\in \mathbb{Z}^{n}_{4^{s}},\forall c\in \overline{\overline{C}} \right\rbrace, \\
	\overline{\overline{\mathscr{C}}}_{i}&=&\left\lbrace c_{0}+c_{1}^{i}\in \mathbb{Z}^{n}_{4^{s}},\forall c\in \overline{\overline{C}}
	\right\rbrace  , i \in \left\lbrace 1,\dots,m-1\right\rbrace, \\
	& \vdots & \\
	\overline{\overline{\mathscr{C}}}_{2^{m-1}-1}&=&\left\lbrace c_{0}+ c_{1}^{i}+\dots+c_{2^{(m-1)}-1}\in \mathbb{Z}^{n}_{4^{s}},\forall c\in \overline{\overline{C}} \right\rbrace.\\
\end{eqnarray*}
Combining Equations \ref{eq5}, \ref{eq7}, \ref{eq8} and \ref{eq9}, we obtain
\begin{equation}\label{eq10}
\mathfrak{C}=\kappa_{1}\left( \bigoplus_{k=0}^{2^{m-1}-1}\eta_{k}\mathscr{C}_{k}\right) \oplus\kappa_{2}\left( \bigoplus_{k=0}^{2^{m-1}-1}\eta_{k}\overline{\mathscr{C}}_{k}\right) \oplus\kappa_{3}\left( \bigoplus_{k=0}^{2^{m-1}-1}\eta_{k}\overline{\overline{\mathscr{C}}}_{k}\right) 
\end{equation}
By Equation \ref{eq10},we have
\begin{cor}\label{cor1}
 For $i = \overline{1,3} $ and $\zeta  \in \left\lbrace 0, i ,ij, \ldots, 2^{m-1}-1 \right\rbrace $, the element $\kappa_{i}\eta_{\zeta}$ is an idempotent of $\mathfrak{R^{s,m}}$
\end{cor} 
\pf
The same as proposition \ref {prop1} 
\qed
\begin{cor}\label{cor999}
Let $\mathfrak{C}$ be a linear code of lenght $n$ over $\mathfrak{R^{s,m}}$. Then $\Phi(\mathfrak{C})=\Psi_{1}\left( \mathcal{C}\right) \otimes \Psi_{2}\left( \overline{\mathcal{C}}\right) \otimes \Psi_{2}\left( \overline{\overline{\mathcal{C}}}\right) $ and $\vert \mathfrak{C}\vert=\vert \Psi_{1}\left( \mathcal{C}\right)\vert \vert \Psi_{2}\left( \overline{\mathcal{C}}\right)\vert\vert\Psi_{2}\left( \overline{\overline{\mathcal{C}}}\right)\vert $.
\end{cor}
 \begin{thm}
 	Let $\mathfrak{C}$ be a linear code of lenght $n$ over $\mathfrak{R^{s,m}}$. Then $\Phi(\mathfrak{C})=\left(\bigotimes_{j=0}^{2^{m-1}-1}\mathscr{C}_{j}\right)  \otimes \left(\bigotimes_{j=0}^{2^{m-1}-1}\overline{\mathscr{C}_{j}}\right)  \otimes \left(\bigotimes_{j=0}^{2^{m-1}-1}\overline{\overline{\mathscr{C}_{j}}}\right)$
 	and $\vert \mathfrak{C}\vert=\vert \mathscr{C}_{0}\vert \ldots \vert\mathscr{C}_{2^{m-1}-1}\vert\vert\overline{\mathscr{C}_{0}}\vert \ldots \vert \overline{\mathscr{C}_{2^{m-1}-1}}\vert \vert\overline{\overline{\mathscr{C}_{0}}}\vert \ldots \vert\overline{\overline{\mathscr{C}_{2^{m-1}-1}}}\vert $.
 \end{thm}
\pf
Let
$
y=(a_{0}^{1},a_{0}^{2},\dots,a_{0}^{n},a_{0}^{1}+ a_{1}^{1i},a_{0}^{2}+ a_{1}^{2i},\dots,a_{0}^{n}+ a_{1}^{ni},\dots,a_{0}^{1}+a_{1}^{1i}+\dots+a_{2^{(m-1)}-1}^{1},
a_{0}^{2}+a_{1}^{2i}+\dots+a_{2^{(m-1)}-1}^{2},\dots,a_{0}^{n}+a_{1}^{ni}+\dots+a_{2^{(m-1)}-1}^{n};b_{0}^{1},b_{0}^{2},\dots,b_{0}^{n},b_{0}^{1}+ b_{1}^{1i},
b_{0}^{2}+ b_{1}^{2i},\dots,b_{0}^{n}+ b_{1}^{ni},\dots,b_{0}^{1}+b_{1}^{1i}+\dots+b_{2^{(m-1)}-1}^{1},
b_{0}^{2}+b_{1}^{2i}+\dots+b_{2^{(m-1)}-1}^{2},\dots,
b_{0}^{n}+b_{1}^{ni}+\dots+b_{2^{(m-1)}-1}^{n}c_{0}^{1},c_{0}^{2},\dots,c_{0}^{n},c_{0}^{1}+ c_{1}^{1i},c_{0}^{2}+ c_{1}^{2i},\dots,c_{0}^{n}+ c_{1}^{ni},\dots,
c_{0}^{1}+c_{1}^{1i}+\dots+c_{2^{(m-1)}-1}^{1},
c_{0}^{2}+c_{1}^{2i}+\dots+c_{2^{(m-1)}-1}^{2},\dots,c_{0}^{n}+c_{1}^{ni}+\dots+c_{2^{(m-1)}-1}^{n})$ in $ \Phi(\mathfrak{C}),
$
then there exists $r=(r_{1},r_{2},\dots,r_{n})\in \mathfrak{C}$, such that $y=\Phi(r)$. So by Equation \ref{eq5}, we have
\begin{equation}
r=\kappa_{1}a+\kappa_{2}b+\kappa_{3}c,
\end{equation}  
where, $a\in\mathcal{C}$, $b\in \overline{\mathcal{C}}$ and $c\in\overline{\overline{\mathcal{C}}}$.

 According to Equations \ref{eq7}, \ref{eq8} and \ref{eq9}, we obtain
 
$\blacktriangleright$ $
a=\eta_{0}(a_{0}^{1},a_{0}^{2},\dots,a_{0}^{n})+\eta_{i}(a_{0}^{1}+a_{1}^{1i},a_{0}^{2}+ a_{1}^{2i},\dots,a_{0}^{n}+ a_{1}^{ni})+\dots+\eta_{_{2^{(m-1)}-1}}(a_{0}^{1}+a_{1}^{1i}+\dots
+a_{2^{(m-1)}-1}^{1},a_{0}^{2}+a_{1}^{2i}
+\dots+a_{2^{(m-1)}-1}^{2},a_{0}^{n}+a_{1}^{ni}
+\dots+a_{2^{(m-1)}-1}^{n}).
$\\
 
$\blacktriangleright$ $
	b=\eta_{0}(b_{0}^{1},b_{0}^{2},\dots,b_{0}^{n})+\eta_{i}(b_{0}^{1}+b_{1}^{1i},b_{0}^{2}+ b_{1}^{2i},\dots,b_{0}^{n}+ b_{1}^{ni})+\dots+\eta_{_{2^{(m-1)}-1}}(b_{0}^{1}+b_{1}^{1i}
	+\dots+b_{2^{(m-1)}-1}^{1},b_{0}^{2}+b_{1}^{2i}
	+\dots+b_{2^{(m-1)}-1}^{2},,b_{0}^{n}+b_{1}^{ni}
	+\dots+b_{2^{(m-1)}-1}^{n}).
$\\
 
$\blacktriangleright$ $
	c=\eta_{0}(c_{0}^{1},c_{0}^{2},\dots,c_{0}^{n})+\eta_{i}(c_{0}^{1}+c_{1}^{1i},c_{0}^{2}+ c_{1}^{2i},\dots,c_{0}^{n}+ c_{1}^{ni})+\dots+\eta_{_{2^{(m-1)}-1}}(c_{0}^{1}+c_{1}^{1i}
	+\dots+c_{2^{(m-1)}-1}^{1},c_{0}^{2}+c_{1}^{2i}
	+\dots+c_{2^{(m-1)}-1}^{2},c_{0}^{n}+c_{1}^{ni}
	+\dots+c_{2^{(m-1)}-1}^{n}).
$

We can sum all of these vectors in the vector $r$ in $\mathfrak{C}$ by, the following way
\begin{eqnarray*}
r&=& \kappa_{1}( \eta_{0}(a_{0}^{1},a_{0}^{2},\dots,a_{0}^{n})+\eta_{i}(a_{0}^{1}+ a_{1}^{1i},a_{0}^{2}+ a_{1}^{2i},\dots,a_{0}^{n}+ a_{1}^{ni})+\eta_{2^{(m-1)}-1} \\
&(& a_{0}^{1}+a_{1}^{1i}+\dots+a_{2^{(m-1)}-1}^{1},a_{0}^{2}+a_{1}^{2i}+\dots+a_{2^{(m-1)}-1}^{2},\dots,a_{0}^{n}+a_{1}^{ni}+\dots+a_{2^{(m-1)}-1}^{n}) )\\
&+& \kappa_{2}( \eta_{0}(b_{0}^{1},b_{0}^{2},\dots,b_{0}^{n})+\eta_{i}(b_{0}^{1}+ b_{1}^{1i},b_{0}^{2}+ b_{1}^{2i},\dots,b_{0}^{n}+ b_{1}^{ni})+\eta_{2^{(m-1)}-1} \\
&(& b_{0}^{1}+b_{1}^{1i}+\dots+b_{2^{(m-1)}-1}^{1},b_{0}^{2}+b_{1}^{2i}+\dots+b_{2^{(m-1)}-1}^{2},\dots,b_{0}^{n}+b_{1}^{ni}+\dots+b_{2^{(m-1)}-1}^{n}) )\\
	&+& \kappa_{3}( \eta_{0}(c_{0}^{1},c_{0}^{2},\dots,c_{0}^{n})+\eta_{i}(c_{0}^{1}+ c_{1}^{1i},c_{0}^{2}+ c_{1}^{2i},\dots,c_{0}^{n}+ c_{1}^{ni})+\eta_{2^{(m-1)}-1} \\
	&(& c_{0}^{1}+c_{1}^{1i}+\dots+c_{2^{(m-1)}-1}^{1},c_{0}^{2}+c_{1}^{2i}+\dots+c_{2^{(m-1)}-1}^{2},\dots,c_{0}^{n}+c_{1}^{ni}+\dots+c_{2^{(m-1)}-1}^{n}) ).
\end{eqnarray*}	
This construction leads to the following,

$
(a_{0}^{1},a_{0}^{2},\dots,a_{0}^{n}) \in  \mathscr{C}_{0},$

$ 
(a_{0}^{1}+ a_{1}^{1i},a_{0}^{2}+ a_{1}^{2i},\dots,a_{0}^{n}+ a_{1}^{ni})\in \mathscr{C}_{i},$

$\vdots$

$(a_{0}^{1}+a_{1}^{1i}+\dots+a_{2^{(m-1)}-1}^{1},a_{0}^{2}+a_{1}^{2i}+\dots+a_{2^{(m-1)}-1}^{2},\dots,a_{0}^{n}+a_{1}^{ni}+\dots+a_{2^{(m-1)}-1}^{n})\in  \mathscr{C}_{{2^{(m-1)}-1}}.
$

The above vectors and Corollary \ref{cor999}, immediately lead to the following result
\begin{eqnarray*}
	&&(a_{0}^{1},a_{0}^{2},\dots,a_{0}^{n},a_{0}^{1}+ a_{1}^{1i},a_{0}^{2}+ a_{1}^{2i},\dots,a_{0}^{n}+ a_{1}^{ni},\dots,a_{0}^{1}+a_{1}^{1i}+\dots+a_{2^{(m-1)}-1}^{1},\\
	&&a_{0}^{2}+a_{1}^{2i}+\dots+a_{2^{(m-1)}-1}^{2},\dots,a_{0}^{n}+a_{1}^{ni}+\dots+a_{2^{(m-1)}-1}^{n})\in \bigotimes_{j=0}^{2^{m-1}-1}\mathscr{C}_{j}=\Psi_{1}\left( \mathcal{C}\right) , 
\end{eqnarray*}
\begin{eqnarray*}
	&&(b_{0}^{1},b_{0}^{2},\dots,b_{0}^{n},b_{0}^{1}+ b_{1}^{1i},b_{0}^{2}+ b_{1}^{2i},\dots,b_{0}^{n}+ b_{1}^{ni},\dots,b_{0}^{1}+b_{1}^{1i}+\dots+b_{2^{(m-1)}-1}^{1},\\
	&&b_{0}^{2}+b_{1}^{2i}+\dots+b_{2^{(m-1)}-1}^{2},\dots,b_{0}^{n}+b_{1}^{ni}+\dots+b_{2^{(m-1)}-1}^{n})\in \bigotimes_{j=0}^{2^{m-1}-1}\overline{\mathscr{C}_{j}}=\Psi_{2}\left( \overline{\mathcal{C}}\right)   
\end{eqnarray*}		
and
\begin{eqnarray*}
	&&(c_{0}^{1},c_{0}^{2},\dots,c_{0}^{n},c_{0}^{1}+ c_{1}^{1i},c_{0}^{2}+ c_{1}^{2i},\dots,c_{0}^{n}+ c_{1}^{ni},\dots,c_{0}^{1}+c_{1}^{1i}+\dots+c_{2^{(m-1)}-1}^{1},\\
	&&c_{0}^{2}+c_{1}^{2i}+\dots+c_{2^{(m-1)}-1}^{2},\dots,c_{0}^{n}+c_{1}^{ni}+\dots+c_{2^{(m-1)}-1}^{n})\in \bigotimes_{j=0}^{2^{m-1}-1}\overline{\overline{\mathscr{C}_{j}}}=\Psi_{2}\left( \overline{\overline{\mathcal{C}}}\right).  
\end{eqnarray*}

So, $y$ is an element in $ \Psi_{1}\left( \mathcal{C}\right) \otimes\Psi_{2}\left( \overline{\mathcal{C}}\right) \otimes\Psi_{2}\left( \overline{\overline{\mathcal{C}}}\right).$   

Conversely, we consider $ y $ defined above an element in $  \Psi_{1}\left( \mathcal{C}\right) \otimes\Psi_{2}\left( \overline{\mathcal{C}}\right) \otimes\Psi_{2}\left( \overline{\overline{\mathcal{C}}}\right),$
with
$
	(a_{0}^{1},a_{0}^{2},\dots,a_{0}^{n})\in \mathscr{C}_{0}, (a_{0}^{1}+ a_{1}^{1i},a_{0}^{2}+ a_{1}^{2i},\dots,a_{0}^{n}+ a_{1}^{ni})\in \mathscr{C}_{i},\dots,
	(a_{0}^{1}+a_{1}^{1i}+\dots+a_{2^{(m-1)}-1}^{1},a_{0}^{2}+a_{1}^{2i}+\dots+a_{2^{(m-1)}-1}^{2},\dots,a_{0}^{n}+a_{1}^{ni}+\dots+a_{2^{(m-1)}-1}^{n})\in \mathscr{C}_{{2^{(m-1)}-1}}.
$

Using Equations \ref{eq6}, \ref{eq7} and \ref{eq8}, we have 
\begin{eqnarray*}
	&&a=\eta_{0}(a_{0}^{1},a_{0}^{2},\dots,a_{0}^{n})+\eta_{i}(a_{0}^{1}+a_{1}^{1i},a_{0}^{2}+ a_{1}^{2i},\dots,a_{0}^{n}+ a_{1}^{ni})+\dots+\eta_{_{2^{(m-1)}-1}}(a_{0}^{1}+a_{1}^{1i}
	\\ &&+\dots
	+ a_{2^{(m-1)}-1}^{1},a_{0}^{2}+a_{1}^{2i}
	+\dots+a_{2^{(m-1)}-1}^{2},\dots,a_{0}^{n}+a_{1}^{ni}+\dots+a_{2^{(m-1)}-1}^{n})\in \mathcal{C},
\end{eqnarray*}
\begin{eqnarray*}
	&&b=\eta_{0}(b_{0}^{1},b_{0}^{2},\dots,b_{0}^{n})+\eta_{i}(b_{0}^{1}+b_{1}^{1i},b_{0}^{2}+ b_{1}^{2i},\dots,b_{0}^{n}+ b_{1}^{ni})+\dots+\eta_{_{2^{(m-1)}-1}}(b_{0}^{1}+b_{1}^{1i}\\
	&+&\dots+b_{2^{(m-1)}-1}^{1},b_{0}^{2}+b_{1}^{2i}
	+\dots+b_{2^{(m-1)}-1}^{2},\dots,b_{0}^{n}+b_{1}^{ni}+\dots+b_{2^{(m-1)}-1}^{n})\in \overline{\mathcal{C}},
\end{eqnarray*}
and
\begin{eqnarray*}
	&&c=\eta_{0}(c_{0}^{1},c_{0}^{2},\dots,c_{0}^{n})+\eta_{i}(c_{0}^{1}+c_{1}^{1i},c_{0}^{2}+ c_{1}^{2i},\dots,c_{0}^{n}+ c_{1}^{ni})+\dots+\eta_{_{2^{(m-1)}-1}}(c_{0}^{1}+c_{1}^{1i}\\
	&+&\dots+c_{2^{(m-1)}-1}^{1},c_{0}^{2}+c_{1}^{2i}
	+\dots+c_{2^{(m-1)}-1}^{2},\dots,c_{0}^{n}+c_{1}^{ni}+\dots+c_{2^{(m-1)}-1}^{n})\in \overline{ \overline{\mathcal{C}}}.
\end{eqnarray*}
 we result, $\kappa_{1}a+\kappa_{2}b+\kappa_{3}c$ in $\mathfrak{C} $. 
 Hence $y=\Phi(\kappa_{1}a+\kappa_{2}b+\kappa_{3}c)\in\Phi(\mathfrak{C})$, so $y\in \Phi(\mathfrak{C})$.
 \qed
\begin{thm}
The generator matrix of $\mathfrak{C}$, a linear code of length $n$ over $\mathfrak{R^{s,m}}$ is 
\begin{eqnarray*}
	\mathcal{G}=\begin{pmatrix}
		G_{1}\\
		G_{2}\\
		G_{3}
	\end{pmatrix},
\end{eqnarray*}
with
\begin{eqnarray*}
	G_{l}=\begin{pmatrix}
		\kappa_{l}\eta_{0}\mathscr{G}_{l0}\\
		\kappa_{l}\eta_{1}\mathscr{G}_{l1}\\
		\vdots\\
		\kappa_{l}\eta_{2^{(m-1)}-1}\mathscr{G}_{l{2^{(m-1)}-1}}
	\end{pmatrix}, \  for \  1\leq l \leq 3.
\end{eqnarray*}
\end{thm}
\pf 
By corollary \ref{cor1}, we obtain
\begin{eqnarray*}
	G=\begin{pmatrix}
	\kappa_{1}\eta_{0}\mathscr{G}_{10}\\ 
	\vdots \\ \kappa_{1}\eta_{2^{(m-1)}-1}\mathscr{G}_{12^{(m-1)}-1}\\
	\kappa_{2}\eta_{0}\mathscr{G}_{20}\\
	\vdots\\
	\kappa_{2}\eta_{2^{(m-1)}-1}\mathscr{G}_{22^{(m-1)}-1}\\
\kappa_{3}\eta_{0}\mathscr{G}_{30}\\
\vdots\\
\kappa_{3}\eta_{2^{(m-1)}-1}\mathscr{G}_{32^{(m-1)}-1}
	\end{pmatrix}
\end{eqnarray*}
 \qed
 \begin{prop}
 If $\mathfrak{C}$ is a linear code of length $n$ over $\mathfrak{R^{s,m}}$ with generator matrix $\mathcal{G}$, then 
 \begin{eqnarray*}
 	\Phi(\mathcal{G})=\begin{pmatrix}
 		&\mathscr{G}_{10}&\mathscr{G}_{10}&\dots&&& \mathscr{G}_{10}\\ 
 		&0&\mathscr{G}_{11}&0&\dots&&\mathscr{G}_{11}\\
 		&\vdots \\
 		&0&0&0&\dots&0&\mathscr{G}_{12^{(m-1)}-1}\\
 		&\mathscr{G}_{20}&\mathscr{G}_{20}&\dots&&& \mathscr{G}_{20}\\ 
 		&0&\mathscr{G}_{21}&0&\dots&&\mathscr{G}_{21}\\
 		&\vdots \\
 		&0&0&0&\dots&0&\mathscr{G}_{22^{(m-1)}-1}\\
 		&\mathscr{G}_{30}&\mathscr{G}_{30}&\dots&&& \mathscr{G}_{30}\\ 
 		&0&\mathscr{G}_{11}&0&\dots&&\mathscr{G}_{31}\\
 		&\vdots \\
 		&0&0&0&\dots&0&\mathscr{G}_{32^{(m-1)}-1}\\
 	\end{pmatrix}
 \end{eqnarray*}
 \end{prop}

\section{Quasi Cyclic codes from Simplex codes over $\mathfrak{R^{s,m}}$ }

Quasi-cyclic codes are a generalization of cyclic codes where every cyclic shift of a codeword is not necessarily a codeword, but only every $d$-th cyclic shift, where $d$ is a divisor of the code length. Quasi-cyclic codes are a natural extension of cyclic codes and they have some interesting properties that make them useful in coding theory.
The quasi cyclic codes has been studied by many researchers.In our work we'll take a look at quasi cyclic codes from simplex codes over $\mathfrak{R^{s,m}}$. We want describe 1-generator,  2-generator until p-generator QC generator matrix.This idea is new and we will discuss it in detail and we will try to review some applications on it.

\subsection{Simplex and MacDonald codes of Type $\alpha$ and $\beta$ over $\mathfrak{R^{s,m}}$ and their Gray Images over $Z_{4^{s}}$ }
In this section we will constructed Simplex and MacDonald codes of Type $\alpha$ and $\beta$ over $\mathfrak{R^{s,m}}$ and their Gray Images over $Z_{4^{s}}$. In \cite{Chatouh}, 
the simplex codes over $R_{q,p,m}$ and  $R_{q}$ of type $\alpha$ and $\beta$ were constructed. 
  We will applicate the same method over our ring $\mathfrak{R^{s,m}}$. We will cite just the results as points without detailed.\\ So we have the simplex codes $S^{\alpha}_{(m,s,k)}$ of length $4^{ms2^{m-1}k}$ and type $\alpha$ over $\mathfrak{R^{s,m}}$ , its Gray images $\phi(S^{\alpha}_{(m,s,k)})=$ $\mathfrak{S}^{\alpha}_{s,k}$ is the concatenation of $4^{ms(2^{m-1})(k+1)}$ simplex codes $\mathfrak{S}^{\alpha}_{s,k}$ over $Z_{4^{s}}$ of length $4^{ms[2^{m-1}(k+1)-1]}$. the similar result about the simplex codes $S^{\beta}_{(m,s,k)}$  of length $\dfrac{4^{ms(2^{m-1})(k-1)+m(s-1)}(4^{mk}-1)}{3}$ and type $\beta$ over $\mathfrak{R^{s,m}}$ have Gray images $\phi(S^{\beta}_{(m,s,k)})=$ $\mathfrak{S}^{\beta}_{s,k}$ wich is the concatenation of $4^{mk[s(2^{m-1}-2)+1]+2m(s-1)}$ simplex codes $\mathfrak{S}^{\beta}_{s,k}$ over $Z_{4^{s}}$ of length \\ 
 $$\dfrac{4^{msk(2^{m-1}-1)+m(s-1)}(4^{mk}-1)}{3}.$$  
		 The MacDonald codes $M^{\alpha}_{s,m,k,u}$ of types $\alpha$ and $\beta$  over $\mathfrak{R^{s,m}}$, for $1\leq u\leq k-1$ is a code over $\mathfrak{R^{s,m}}$ of length $4^{ms2^{m-1}k}-4^{ms2^{m-1}u}$. Its gray images$\phi(M^{\alpha}_{s,m,k,u})$ is the concatenation of $$\dfrac{4^{ms2^{m-1}(k-1)-ms}-4^{ms2^{m-1}(u-1)-ms}}{4^{msk}-4^{msu}}$$ MacDonald codes $\mathfrak{M}^{\alpha}_{s,k}$  over $Z_{4^{s}}$ of length $$4^{m[s2^{m-1}(k+1)-1]}-4^{m[s2^{m-1}(u+1)-1]}.$$ The MacDonald codes $M^{\beta}_{s,m,k,u}$ is a code over $\mathfrak{R^{s,m}}$ of length $$\dfrac{2^{m[s(4^{m-1}-1)(k-1)+(s-1)]}(4^{mk}-1)-4^{m[s(2^{m-1}-1)(u-1)+(s-1)]}(4^{mu}-1)}{3}.$$ Its gray images $\phi(M^{\beta}_{s,m,k,u})$ is the concatenation of  $$\dfrac{4^{msk(2^{m-1}-1)+m(s-1)}(4^{mk}-1)-4^{msk(2^{m-1}-1)+m(s-1)}(4^{mu}-1)}{4^{ms(k-1)}(4^{mk}-1)-4^{ms(u-1)}(4^{uk}-1)}$$ of MacDonald codes $\mathfrak{M}^{\beta}_{s,k}$ over $Z_{4^{s}}$ of length  $$\dfrac{4^{m[sk(2^{m-1}-1)+(s-1)]}(4^{mk}-1)-4^{m[sk(2^{m-1}-1)+(s-1)]}(4^{mu}-1)}{3}.$$
		 	
 \section{Conclusion}
The subject discussed is a generalization of a ring $\mathfrak{R^{s,m}}$ and its properties, including cyclic codes and quasi-cyclic codes over$\mathfrak{R^{s,m}}$. Cyclic codes are a special type of linear  code where every cyclic shift of a codeword is also a codeword. Quasi-cyclic codes are a generalization of cyclic codes where every $d$-th cyclic shift of a codeword is a codeword, where $d$ is a divisor of the code length.


In conclusion, the generalization of a ring $\mathfrak{R^{s,m}}$ provides a rich mathematical structure for the study of cyclic and quasi-cyclic codes, and has practical applications in error correction and information transmission. Simplex codes over R are a particular example of quasi-cyclic codes with interesting algebraic properties, and they can be constructed using the simplex algorithm over a division ring.

\end{document}